\documentclass{desyproc}

\begin{document}
\title{Shining Light on Dark Energy and Modifications of Gravity}

\author{{\slshape Clare Burrage}\\[1ex]
School of Physics and Astronomy, University of Nottingham, Nottingham NG7 2RD, UK}


\desyproc{DESY-PROC-2012-04}
\acronym{Patras 2012} 
\doi  

\maketitle

\begin{abstract}
Many theories of dark energy and modified gravity give rise to scalar fields that couple derivatively to the energy momentum tensor of matter. This is known as disformal coupling. I will show that laboratory searches for axions are ideally suited to search for and constrain disformal scalar fields.
\end{abstract}

\section{Introduction}

We now have conclusive evidence that the expansion of the universe is accelerating.  Explanations that go beyond a cosmological constant, and the associated extreme fine tuning problem, require either the introduction of unusual matter fields or the modification of gravity.  In practice (and excepting extreme environments such as black holes) both of these explanations almost always introduce a new scalar field either directly or indirectly.    As the cause of the acceleration of the expansion of the universe must be coherent across the whole of the observable universe today any scalar fields associated with an explanation  of the acceleration  must be extremely light with masses $m_{\phi}\sim H_0\sim 10^{-33}\mbox{ eV}$.

Interactions between these scalar fields and matter  must exist; there are no known symmetries or principles that can forbid all couplings.  So how do these new scalar fields interact with the fields of the Standard Model?  In these proceedings I will discuss the most general way a scalar field can interact with Standard Model matter, and the signatures of these interactions that could be detected by high precision photon experiments.

\section{The most general  interactions with matter}
We expect that the scalar field interacts with all matter fields in the same way, therefore the scalar field defines a new metric which is felt by the matter fields.  The  Lagrangian  describing interactions between the scalar and matter is:  $\mathcal{L}\supset \tilde{g}_{\mu\nu}T^{\mu\nu},$
where $\tilde{g}_{\mu\nu}$ is a metric which depends on the scalar field $\phi$ and the Einstein metric $g_{\mu\nu}$. The energy momentum tensor of all of the Standard Model fields in our theory is  $T^{\mu\nu}$.  Bekenstein proved, \cite{Bekenstein:1992pj}, that the most general form for $\tilde{g}_{\mu\nu}$ which respects causality and the weak equivalence principle is 
\begin{equation}
\tilde{g}_{\mu\nu}=A(\phi,X)g_{\mu\nu} +B(\phi,X)\partial_{\mu} \phi\partial_{\nu}\phi\;,
\label{eq:bek}
\end{equation}
where $X=-(\partial \phi)^2/2$.  The term containing $A(\phi,X)$ is known as the conformal term, as when only this term is present the metric $\tilde{g}_{\mu\nu}$ is a conformal rescaling of the metric $g_{\mu\nu}$.  The term containing $B(\phi,X)$ is known, in contrast, as the disformal term.  If the scalar field couples to matter fields through a conformal term in the metric then the scalar field will give rise to a fifth force felt by matter fields. When only the disformal term is present  the scalar field is not sourced by a static non-relativistic distribution of matter, and so no fifth forces arise.

\subsection{Disformal couplings from massive gravity}
The conformal coupling has been much studied, but the disformal term is less familiar.  It arises in a number of  contexts, and of particular interest  is that disformal couplings appear in theories of massive gravity. The question of how to give the graviton a mass without causing ghost degrees of freedom to occur was only answered recently by de Rham, Gabadadze and Tolley \cite{deRham:2010ik,deRham:2010kj}. The mass of the graviton must be $m \lesssim H_0$, in order to match the observed cosmology, and if gravity starts to become weaker on the largest scales in the universe due to a Yukawa suppression this could explain the observed acceleration of the expansion of the universe. Care is needed when studying massive gravity however because it suffers from problems of strong coupling \cite{Burrage:2012ja}.

  A massive spin-two field can be decomposed into massless degrees of freedom;  one helicity-two mode, two helicity-one modes and one helicity-zero mode.  At low energies the vector degrees of freedom decouple from all other fields and so can be consistently set to zero\footnote{The existence of additional helicity-one degrees of freedom might also give rise to interesting signals in high-precision photons experiments as couplings to other fields do exist in the full theory.  However the interactions are not fully understood at present.}.  In  massive gravity matter fields couple to a metric
\begin{equation}
g_{\mu\nu}=\left(1+\frac{\phi}{M_P}\right)\eta_{\mu\nu}+\frac{u}{M_P^2 m^2}\partial_{\mu}\phi \partial_{\nu}\phi\;,
\end{equation}
where $m$ is the mass of the graviton,  $u$ is a dimensionless constant of order one and $\eta_{\mu\nu}$ and $\phi$ are the helicity-two and helicity-one parts of the graviton respectively.  The strength of the disformal terms is controlled by the scale  $M\approx (M_P m)^{1/2}\approx 10^{-2}\mbox{ eV}$.  This is a surprisingly low scale for new physics that makes looking for disformal couplings an attractive possibility to detect massive gravity.

The scalar field arising for massive gravity has higher order derivative self interactions in its Lagrangian.  These mean that while our estimate of the scale $M$ controlling the disformal terms is correct in vacuum the scale varies depending on the background scalar field configuration.  A massive source object, such as the Earth, gives rise to a non-trivial scalar field profile in massive gravity.  The presence of the non-linear scalar self couplings means that the coupling constants of the scalar field become dependent on the background scalar field profile.  How this occurs is known as the Vainshtein effect and is discussed in more detail in \cite{deRham:2012az,Burrage:2012ja}.  The coupling constants of the scalar field in massive gravity get rescaled in the following way: $M_P\rightarrow  Z(\phi_0) M_P$ and $M=(M_P m)^{1/2}  \rightarrow Z^{1/2}(\phi_0) M$,
where $\phi_0$ is the background field configuration and $Z(\phi_0)\gg 1$.  This raises the scale at which the field couples to matter.  In particular the scale of the disformal coupling that we found above $M\sim 10^{-2}\mbox{ eV}$ should be thought of as a lower bound on the expected value of $M$ in an experimental environment at the surface of the Earth.

\section{Interactions with photons}
If only a conformal coupling is present $\tilde{g}_{\mu\nu}=A(\phi)g_{\mu\nu}$ there is no coupling to the kinetic terms of gauge bosons at tree level.  
However the  Standard Model is not conformally invariant, and the conformal anomaly means that a coupling between the scalar field and photons will always be generated \cite{Brax:2010uq}.  The scale of this coupling is not determined by this calculation and can only be fixed by experimental observations.

As mentioned before the conformal coupling has been much studied in the literature, whereas the disformal coupling has been largely neglected.  Therefore to understand the phenomenology of the disformal term we focus on a metric that is a simple form of Bekenstein's metric in Equation (\ref{eq:bek})
\begin{equation}
\tilde{g}_{\mu\nu}=\left(1+\frac{\phi}{\tilde{\Lambda}}\right)g_{\mu\nu}+\frac{2}{M^4}\partial_{\mu}\phi\partial_{\nu}\phi \;.
\end{equation}
Then the Lagrangian describing the scalar field and photon is 
\begin{equation}
\mathcal{L}_{\phi,\gamma}=-\frac{1}{2}(\partial\phi)^2 -V(\phi) -\frac{1}{4}F^2 -\frac{\phi}{\Lambda}F^2 -\frac{1}{M^4}\partial_{\mu}\phi\partial_{\nu}\phi \left[\frac{1}{4}g^{\mu\nu}F^2 -F^{\mu}_{\;\; \alpha}F^{\nu\alpha}\right]\;,
\label{eq:lagphotscal}
\end{equation}
where, as usual, $F_{\mu\nu}=\partial_{\mu}A_{\nu}-\partial_{\nu}A_{\mu}$.

The term controlled by the scale $\Lambda$ is the coupling between scalars and photons studied for axion-like particles, that arises because of the conformal term in the metric.  Couplings between scalars and photons modify the propagation of photons through a magnetic field causing both changes in polarization and non-conservation of photon number.  In a coherent magnetic field of strength $B$ the relevance of the disformal couplings are controlled by the dimensionless combination $b=B/M^2$. 

The evolution of the scalars and photons through a constant magnetic field oriented perpendicular  to the direction of motion is given by a set of coupled differential equations.  These can be diagonalized and solved to give the probability of a photon oscillating into a scalar (or vice versa): $
P_{\gamma\leftrightarrow\phi}=\sin^22\vartheta\sin^2\lambda\omega x,$
where
\begin{equation}
\tan 2\vartheta =\frac{4 B}{\Lambda\omega} \sqrt{\frac{1+b^2}{1-a^2}}\left(\frac{m^2}{2\omega^2}-b^2\right)^{-1}\;,
\end{equation}
\begin{equation}
\lambda_{\pm}=-\lambda(\cos2\vartheta \mp 1)\;,\;\;\;\;\;\lambda =\frac{1}{2(1+b^2)}\left|\frac{m^2}{2\omega^2} -b^2\right|(1+\tan^2 2\vartheta)^{1/2}\;.
\label{eq:eigenvalues}
\end{equation}
These reduce to the familiar expressions for  axion-like particles when $b=0$.  As no signs of mixing between scalars and photons have yet been seen in the laboratory these experiments will only be sensitive to the regime where the mixing is weak $\vartheta \ll1$.  In the weak mixing regime
\begin{equation}
P\approx \left(\frac{4B}{\Lambda \omega}\right)^2 \frac{1+b^2}{(\frac{m^2}{2\omega^2}-b^2)^2}\sin^2\left[\frac{(\frac{m^2}{2\omega^2}-b^2)\omega x}{2(1+b^2)}\right]\;.
\end{equation}

\begin{figure}[ht]
\centerline{\includegraphics[width=0.35\textwidth]{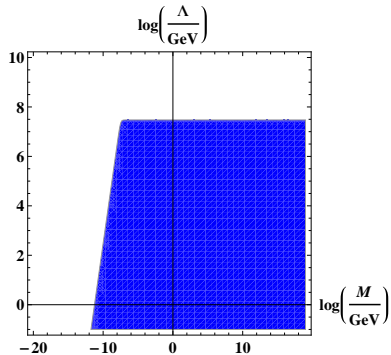}}
\caption{Constraints on the conformal ($\Lambda$) and disformal ($M$) coupling strengths given by the ALPS experiment in the small mass limit (we have set $m\sim 10^{-21} \mbox{ eV}$). The shaded region is excluded.}\label{Fig:MV}
\label{sec:figures}
\end{figure}

In the small mass limit that we are interested in we see that there are three regimes of behaviour: $b\ll m/\omega\ll 1$ where the disformal terms are negligible and the field behaves just like an axion-like particle, and two regimes where the disformal terms are important $m/\omega \ll b \ll 1$ and $1\ll b$.  When the disformal terms are important the probability of mixing is always less than the probability of mixing in the axion-like particle case with the same value of $\Lambda$.  So we expect experiments only to be sensitive to the initial onset of the disformal terms, this is the  regime $m/\omega \ll b \ll 1$.  Then  the effects of the disformal coupling can be encoded in  an effective mass for the scalar field $m_{\rm eff}^2 =|m^2-2\omega^2 b^2|$.
  The constraints from the ALPS experiment \cite{Ehret:2010mh} are plotted in Figure \ref{Fig:MV}. We see that for $\Lambda\lesssim 10^7 \mbox{ GeV}$ massive gravity strength disformal couplings are already excluded.

\section{Conclusions}
Dark energy and theories of modified gravity typically introduce new scalar degrees of freedom.  The most general way that these scalar fields can couple to matter includes both conformal and disformal terms.  These disformal couplings have been largely neglected to date, but high-precision photon experiments can  constrain them. The ALPS experiment at DESY is already  probing the regime of parameter space that is interesting for massive gravity models.

\section*{Acknowledgments}
This work was  done in collaboration with Philippe Brax and Anne-Christine Davis and published as \cite{Brax:2012ie}.
CB is supported by a University of Nottingham Anne McLaren Fellowship.



\begin{footnotesize}

\end{footnotesize}



\begin{thebibliography}{99}


\bibitem{Bekenstein:1992pj}
  J.~D.~Bekenstein,
  Phys.\ Rev.\ D {\bf 48} (1993) 3641

\bibitem{deRham:2010ik}
  C.~de Rham and G.~Gabadadze,
  Phys.\ Rev.\ D {\bf 82} (2010) 044020

\bibitem{deRham:2010kj}
  C.~de Rham, G.~Gabadadze and A.~J.~Tolley,
  Phys.\ Rev.\ Lett.\  {\bf 106} (2011) 231101
\bibitem{deRham:2012az}
  C.~de Rham,
  Comptes Rendus Physique {\bf 13} (2012) 666

\bibitem{Burrage:2012ja}
  C.~Burrage, N.~Kaloper and A.~Padilla,
  arXiv:1211.6001 [hep-th].

\bibitem{Brax:2010uq}
  P.~Brax, C.~Burrage, A.~-C.~Davis, D.~Seery and A.~Weltman,
  Phys.\ Lett.\ B {\bf 699} (2011) 5
\bibitem{Ehret:2010mh}
  K.~Ehret, M.~Frede, S.~Ghazaryan, M.~Hildebrandt, E.~-A.~Knabbe, D.~Kracht, A.~Lindner and J.~List {\it et al.},
  Phys.\ Lett.\ B {\bf 689} (2010) 149


\bibitem{Brax:2012ie}
  P.~Brax, C.~Burrage and A.~-C.~Davis,
  JCAP {\bf 1210} (2012) 016





\end{thebibliography}
\end{document}